\documentclass[fleqn,12pt,twoside]{article}
\usepackage{graphicx}
\usepackage{amsmath}
\usepackage{amsfonts}
\usepackage{dcolumn}

\newcommand{\Sh}{Schr\"odinger{ }}

\newcommand{\mca}[1]{\mathcal{#1}}
\newcommand{\bscal}[1]{\bs{\mathcal{#1}}}
\newcommand{\bs}[1]{\boldsymbol{#1}}
\newcommand{\mb}[1]{\mathbb{#1}}
\newcommand{\Fig}[1]{Fig. \ref{#1}}
\newcommand{\braket}[3]{\left\langle #1 \left\vert #2
            \right\vert #3 \right\rangle}

\newcommand{\brak}[2]{\left\langle #1 \left\vert
             #2 \right. \right\rangle}
\newcommand{\ket}[1]{\left\vert #1 \right\rangle}
\newcommand{\fuve}[1]{\left(\bs{#1}\right)}
\newcommand{\fues}[1]{\left(#1\right)}

\newcommand{\yav}[1]{\left[#1\right]}

\newcommand{\llal}[1]{\left\{#1\right.}
\newcommand{\abs}[1]{\left\vert#1\right\vert}
\newcommand{\num}{\nonumber}
\newcommand{\non}{\nonumber \\}
\newcommand{\delt}[1]{\delta\left(#1\right)}

\newcommand{\Eq}[1]{Eq. (\ref{#1})}
\newcommand{\Eqs}[1]{Eqs. (\ref{#1})}
\newcommand{\rep}[1]{(\ref{#1})}

\newcommand{\nas}[1]{\nabla_{\bs{#1}}}
\newcommand{\sa}{_{\alpha}}

\title{Hofstadter spectrum in electric and magnetic fields. }   
\author{Alejandro Kunold$^1$, Manuel Torres$^2$\\
$^1$Departamento de Ciencias B\'asicas,\\
 Universidad Aut\'onoma Metropolitana-Azcapotzalco,\\
Av. San Pablo 180,  M\'exico D. F. 02200, M\'exico,\\
Email:akb@correo.azc.uam.mx\\ 
$^2$Instituto de F\'{\i}sica,\\
Universidad Nacional Aut\'onoma de M\'exico,\\
Apartado Postal 20-364,  M\'exico D.F. 01000,  M\'exico,\\
Email:torres@fisica.unam.mx}

\begin{document}  

\maketitle
\begin{abstract}
The  problem  of  Bloch electrons  in two dimensions subject to  magnetic and
intense electric fields is investigated.
Magnetic  translations, electric evolution  and energy translation operators are
used to specify the solutions of the Schr\"odinger
equation. For rational values of  the  magnetic flux quanta per unit cell and
commensurate  orientations of the electric field relative to the original lattice,
an extended superlattice can be  defined and a  complete set  of mutually commuting
space-time symmetry operators is obtained.  Dynamics of the system is  governed
by   a finite difference  equation that exactly includes the  effects of:
an arbitrary periodic potential,  an electric field orientated in a commensurable
direction of the lattice,  and  coupling between Landau levels.
A  weak  periodic potential broadens  each Landau level in a series of minibands,
separated by the corresponding minigaps. The addition of  the  electric field induces a series
of avoided and exact  crossing  of the quasienergies, for sufficiently strong electric field 
the spectrum evolves into  equally spaced discreet levels,
in this ``magnetic Stark ladder"  the energy separation is an integer multiple
of  $ h E / a B $, with $a$ the lattice parameter. 
\end{abstract}
\maketitle 

\section{Introduction}\label{intro}

The problem of electrons moving under the  simultaneous influence of a periodic
potential and  a magnetic field has been discussed by many
authors \cite{Peierls1,Harper1,Zak1,Azbel1,Rauh1,Dana1,Harper3,Zak2}; the 
spectrum displays an amazing complexity including various kinds of
scaling and a Cantor set structure \cite{Hofta1}. Some of these results
had been used in order    to disclose the topological structure of  the Hall conductance within the
linear response theory \cite{Thou1,Thou2,Khomo1}.
The addition of an electric field leads to  interesting new phenomena 
that makes  its analysis worthwhile.  For example, it has been suggested that as the strength 
of the electric field increases  the longitudinal quasi-momentum is quantized leading  to the appearance
of a ``magnetic Stark ladder", in which the bands are replaced by a series of quasi discreet
levels \cite{Kunold1}. The quantization of the spectrum  also affects the  the  localization properties
of electron wave packets   \cite{Brito1,Torres1}.

We consider the problem of an electron moving in a two-dimensional
lattice in the presence of applied  magnetic and electric  fields. We refer to this as the electric-magnetic
Bloch problem (EMB).
The corresponding magnetic Bloch system  (MB) has a long and rich
history.
An early important contribution was made by Peierls \cite{Peierls1}
who  suggested the substitution of the  Bloch index $\bs k$ by the
operator $\left(\bs p - e \bs A \right)$ in the $B = 0$ dispersion
relation  $\epsilon(\bs k)$, which is then treated as an effective
single-band Hamiltonian.
The symmetries of the MB problem were analyzed by Zak \cite{Zak1,Dana1,Zak2},
who worked out the representation
theory of the magnetic-translations group. The renowned Harper equation
was derived assuming a tight-binding approximation in which  the magnetic field
acts as a perturbation that splits the Bloch bands  \cite{Harper1}.
Rauh derived a dual Harper equation in the opposite limit of intense magnetic
field \cite{Rauh1},
here the periodic potential acts as a perturbation that broadens the Landau levels,
in this case the Harper equation takes the form  
\begin{equation}\label{harp0}
c_{m-1}+2\cos\fues{2\pi \sigma m + \kappa} c_{m}
+ c_{m+1}=\varepsilon c_{m}.
\end{equation}
where  $\sigma = 1/\phi$ is the  inverse of magnetic flux $\phi$ in a cell
in units of $h/e$.  The studies
of the butterfly spectrum by  Hofstadter and
others   have since created an  unceasing  interest in the problem
because of the beautiful self-similar structure of the
butterfly spectrum \cite{Hofta1,Claro1}.  Remarkably, an experimental realization of the
Hofstadter butterfly  was  achieved  considering  the transmission of
microwaves \cite{Kuhl1} and
acoustic waves  \cite{Richoux1} through an array of scatterers.
However,  the study of these phenomena in electronic transport has only recently  become experimentally accessible
with the developments in the fabrication of antidot
arrays in lateral superlattices by ion beam and atomic force lithography  \cite{Esslin3,Klit4}. For example,   
a signal  of the Hofstadter Butterfly spectrum through
the measurement of the magnetoresistance and Hall conductance in
artificial arrays of anti quantum dots has just been reported \cite{Klit2}.
It is also worthwhile  to point out  that Aubry and Andre \cite{Aubry1}  utilized
 Eq. \ref{harp0} to model a one dimensional system that develops  a metal-insulator transition; this transition, 
arises from the competing effects of two potentials terms,  which are periodic but incommensurate with one another.

The symmetries of the EMB problem were analyzed by Ashby and Miller,
who constructed the group of the electric-magnetic translation
operators, and worked out their irreducible representations \cite{Ashby1}.
The properties of the electric-magnetic operators were utilized in order
to derive a finite difference equation that governs the dynamics of the
EMB problem when  the coupling between Landau levels can be
neglected \cite{Kunold1}. In this paper we shall derive the
equation that applies under most general conditions.
Magnetic  translations, electric evolution and energy translation
operators are used to specify the solutions of the Schr\"odinger
equation, commensurability conditions must be implemented in order
to obtain a set of mutually commuting space-time symmetry
operators. In addition to the broadening of the
Landau levels produced  by the  periodic potential, the 
electric field induces a series
of avoided and exact level crossing   that yields a
generalized Hofstadter spectrum, with an interesting 
pattern that arises out of the competition between 
by the lattice potential and the external $\bs E$ and $\bs B$ fields. 
For sufficiently strong electric field 
the spectrum evolves into  equally spaced discreet levels,
in this ``magnetic Stark ladder"  the energy separation is an integer multiple
of  $ h E / a B $, with $a$ the lattice parameter.

The paper is organized as follows. In the next section we present
the model that describes the EMB problem  and  construct its symmetry
operators. In Section \ref{wave} we describe the commensurability 
conditions required to have simultaneously commuting operators, they
are exploited in order to construct  an appropriated wave function basis.
We derive an effective equation \rep{pro3}, in which the
``evolution" is determined by  a differential equation  with respect to
the longitudinal pseudomomentum.
The derivation of the finite difference equation that governs the
dynamics of the system is presented in Section \ref{harper}.
Results for the  energy spectrum are presented, and are also
discussed from the perspective of the adiabatic approximation. 
The last section contains a summary of our main results. 

\section{The electric-magnetic Bloch problem}\label{embp}
\subsection{The model}\label{model}
  
Let us consider the  motion of an electron in a two-dimensional
periodic potential $V$,  subject  to a uniform magnetic field $B$
perpendicular  to the plane and to a constant electric field $\bs E$,
lying on the plane according to  $\bs E =  E (\cos \theta, \sin \theta)$
with $\theta$ the angle between  $\bs E$ and the lattice 
$x-$axis. The dynamics of the electron is governed by a time-dependent
Schr\"{o}dinger equation that for convenience is written as 
\begin{equation}\label{shr1}
S\ket{\psi}=
\yav{\frac{1}{2 m^*}\fues{\Pi_x^2+\Pi_y^2}+V-\Pi_0}
\ket{\psi}=0,
\end{equation}
here  $m^*$ is the effective  electron mass,  $ \Pi_{\mu}=p_{\mu}+eA_{\mu}$, 
with  $p_{\mu}= (i\hbar \partial/\partial t, - i\hbar \nabla )$.
Throughout the paper energy
and  lengths are  measured in units of 
$ \hbar\omega_c=\frac{\hbar e B}{m^*},$ and $ \ell_0=\sqrt{\frac{\hbar}{e B}}$,
respectively, where  $\omega_c$ is the cyclotron frequency and $ \ell_0$
the magnetic length.  So, unless specified, we set
$\hbar=e=m^*=1$; although $B$ can also be omitted from the expressions,
we find convenient to  explicitly display it.   Covariant notation will
be used  to simplify the expressions, $e.g.$
$x_{\mu}=\fues{t,\bs{x}}=\fues{t,x,y}$.

Equation  (\ref{shr1}) can be considered as an eigenvalue equation for the operator 
$S$ with eigenvalue $0$.  The gauge potential is written  in an arbitrary gauge,
it includes two gauge parameters  $\alpha$, and  $\beta$,
the final physical results should be of course, independent of the gauge. 
Hence the components of the gauge potentials are written as 
\begin{align}
A_0&=  \fues{ \beta- \frac{1}{2} } \, \bs{x} \cdot \bs{E}
,\non
A_x&=-\fues{\beta +\frac{1}{2}}E_xt
+\fues{\alpha -\frac{1}{2}} B y,\label{pve0}\\
A_y&=-\fues{\beta +\frac{1}{2}}E_yt
+\fues{\alpha +\frac{1}{2}} B x \, . \num
\end{align}
The symmetrical gauge is recovered for $\alpha = 0$, whereas the Landau
gauge corresponds to the selection
$\alpha = 1/2$. If $ \beta= 1/2$ all the electric field contribution appears
in the vector potential, instead  if 
$ \beta= - 1/2$ it lies in  the scalar potential and the \Sh equation becomes time 
independent, however even in this case a time dependence will slip  into the problem
through the symmetry operators.   A general two-dimensional periodic potential
can be represented in terms of  its Fourier decomposition
\begin{equation}\label{pot0}
V\fues{x,y}=\sum_{r,s}v_{rs}
\exp\fues{i\frac{2\pi  r x}{a}+i\frac{2\pi s y}{a}} \, . 
\end{equation}
For specific numerical results we shall use  the potential
\begin{equation}\label{pot1}
V\fues{x,y}=U_0\yav{\cos\fues{\frac{2\pi x}{a}}
+\lambda\cos\fues{\frac{2\pi y}{a}}}.
\end{equation}
$\lambda$ is a parameter that can be varied in order to have an anisotropic
lattice; $\lambda=1$ corresponds to the isotropic limit. 

\subsection{Electric evolution and magnetic  translations}\label{magtrans}

Let $(t, \bs x) \to (t + \tau , \bs x + \bs R)$  be
a uniform translation in space and time, where $\tau$ is an
arbitrary time and $\bs R$ is a lattice vector.
The classical equations  of  motion remain  invariant under these
transformations; whereas  the  Schr\"{o}dinger equation does not,
the reason being the space and time dependence of the gauge potentials.
Nevertheless,  quantum dynamics  of the system  remain invariant under
the combined  action of  space-time translations  and  gauge
transformations. The electric and magnetic translation 
operators are defined as
\begin{equation} \label{eqmod3}
T_0(\tau) = \exp{ (- i\tau {\cal O}_0)}  \, , \qquad
 T_j (a) =   \exp{(ia \,  {\cal O}_j )}  \, , 
\end{equation}
with  $j = x,y$ and the  electric-magnetic symmetry generators  are written as
covariant derivatives $ {\cal O}_\mu = p_\mu +   \Lambda_\mu$,  
with the components of the dual gauge potentials $ \Lambda_\mu $ given by
\begin{align}\label{dualgauge}
 \Lambda_0&= A_0 +  \bs x \cdot \bs E \, ,
\non
 \Lambda_x&= A_x + B y + E_x t    \,,   \qquad 
 \Lambda_y  = A_y  - B x + E_y t  \, .  
\end{align}
It is straightforward to prove that  the operators in (\ref{eqmod3})
  are indeed symmetries of the \Sh equation; 
they commute with the operator $S$  in Eq. (\ref{shr1}).   Similar expressions
for the electric-magnetic operators were given by Ashby and Miller \cite{Ashby1},
however their definition  included  simultaneous  space and time translations;
we deemed it  more convenient  to separate the effect of the time evolution
generated by  the    $T_0 $ to that of the space translations generated by
$ T_j$. The following commutators can be worked out 
\begin{align}\label{conmu1}
 \left[\Pi_0,\Pi_j \right] = & -iE_j  \, ,  \qquad \qquad
 \left[\Pi_1,\Pi_2 \right] = -iB \, , 
\nonumber \\
\left[ {\cal O}_0,  {\cal O}_j\right] = &  \,\, iE_j \, ,
\qquad \qquad \,\,\,\,\,
\left[ {\cal O}_1, {\cal O}_2 \right] =iB \, , \nonumber \\
\left[\Pi_\mu, {\cal O}_\nu \right] = & \,\, 0  \, .
\end{align}
Notice that the electric-magnetic generators   ${\cal O}_\mu$ have been defined
in such a form that they commute with all the velocity operators   $\Pi_\nu $.    
The Schr\"{o}dinger equation and the symmetry operators are
expressed in terms of covariant derivatives
$\Pi_\mu$ and ${\cal O}_\mu$, respectively.
A dual situation in which the roles of $\Pi_\mu$ and ${\cal O}_\mu$
are interchanged can  be considered. The dual problem corresponds
to a simultaneous reverse in the directions of $B$ and $\bs E$.
We notice that the commutators in  the second line of Eq. (\ref{conmu1})
are part  of the  Lie  algebra of the EM-Galilean two dimensional
group \cite{Hadji1,Hadji2}. This group is obtained when  the usual rotation
and  boost operators of  the planar-Galilean group  are replaced  by their
electric-magnetic  generalization in which the operators  are enlarged by
the effect of a gauge transformation. 

 \section{Electric-magnetic Bloch Functions}\label{wave}
\subsection{Commensurability conditions}\label{commen}

In order to construct a complete base that  expands the wave function we 
require the symmetry operators to  commute with each other. However we have
\begin{equation}\label{nco1}
T_{\mu}T_{\nu}=
e^{-c^{\mu}c^{\nu}\yav{\mca{O}_{\mu},\mca{O}_{\nu}}} \, 
T_{\nu}T_{\mu} \, ,
\end{equation}
where $c^0=-\tau$, and $c^1=c^2=a$. 
A set of simultaneously commuting symmetry operator can be found if appropriated
commensurate conditions  are imposed, we follow a   three   step method to find them:
\begin{enumerate}
\item First  we consider a  frame rotated  at  angle $\theta$,   with axis along
the longitudinal and transverse direction relative to the electric field.
An orthonormal basis for   this frame is  given by
$\bs{e}_L = (cos \theta, sin \theta)$,  $\bs{e}_T = (-  sin \theta, cos \theta)$
and $\bs{e}_3=\bs{e}_L \times\bs{e}_T$. The electric field is parallel to
$\bs{e}_L$ and the magnetic field points along $\bs{e}_3$. 
We assume a   particular orientation of  the electric field,
for which  the following condition holds
\begin{equation}\label{rho1}
\tan \theta=\frac{E_y}{E_x}=\frac{m_2}{m_1} \, , 
\end{equation}
where  $m_1$ and $m_2$   are relatively prime integers.
This condition insures that spatial periodicity is also found
both along the transverse and the  longitudinal directions.
Hence, we define a rotated lattice spanned by the longitudinal
$\bs b_L = b \bs{e}_L $ and transverse  $\bs b_T = b \bs{e}_T$
vectors, where $b=a\sqrt{m_1^2 +  m_2^2}$.
The  spatial components of the symmetry generator $\bscal O$ are
projected along the longitudinal and transverse directions:
$\mca{O}_L = \bs{e}_L \cdot \bscal{O}$ and
$\mca{O}_T  = \bs{e}_T \cdot \bscal{O}$.
It is readily  verified that $ \yav{\mca{O}_0,\mca{O}_T}=0$.

\item  For the rotated lattice,  we regard  the  number of flux quanta per
unit cell  to be a rational  number  $p/q$, that is 
\begin{equation}\label{sig1}
\phi \equiv \frac{1 }{ \sigma}=\frac{B \, b^2}{ 2\pi }=\frac{p}{q}    \, . 
\end{equation}
We can then define a extended superlattice. A rectangle made of $q$
adjacent lattice cells of side $b$ contains an integer number of  flux quanta.
The basis vectors of the superlattice are chosen as $q \bs b_L$ and $\bs b_T$.
Under these conditions the longitudinal and transverse magnetic translations
$T_L \fues{qb}=\exp{(iq b {\cal O}_L)}$ and
$T_T\fues{b}= \exp{( ib {\cal O}_T)}$
define commuting symmetries under displacements $q\bs b_L$ and $\bs b_T$.  
Henceforth we shall either use the subindex $(L,T)$ or $(1,2)$
to label the longitudinal and transverse directions. 

\item We observe that  $T_0$ and  $T_L(q b)$ commute with 
$T_T$.  Yet they fail to  commute with each other: 
\begin{equation}\label{tau11}
T_0\fues{\tau }T_L \fues{qb}=
e^{-iqb\tau E}
T_L\fues{qb}T_0\fues{\tau}.
\end{equation}
However  the  operators $T_0$ and  $T_L(q b)$  will  commute with one another
by  restricting   time,   in the   evolution operator,  to  discreet values
with period 
\begin{align}
\tau&=n\tau_0,& n&\in \mathbb{Z},&
\tau_0&=\frac{2\pi}{qbE}=
\frac{1}{p}\fues{\frac{b}{v_D}}\label{tau1} \, , 
\end{align}
where the drift velocity is $v_D =  {E / B}$  and  we utilized Eq. (\ref{sig1}) to
write the second equality.
$b/v_D$ is the period of time it takes an electron with  drift velocity $v_D$ to
travel between lattice points. \Eq{tau1} can be interpreted as the condition that
the ratio of  two energy scales is an integer: as we shall see ($\sigma bE$) is
the magnetic-Stark ladder spacing, whereas $(2 \pi v_D /b)$ is the quasienergy 
Brillouin width;  hence Eq. (\ref{tau1}) represents  the ratio of these two quantities.
\end{enumerate}

We henceforth consider that the three conditions (\ref{rho1}),  (\ref{sig1}), and
(\ref{tau1}) hold simultaneously. In this case the three EM operators:
the electric evolution ${\cal T}_0 \equiv T_0(\tau_0)$, and the magnetic translations
${\cal T}_L \equiv T_L(q b)$, and ${\cal T}_T \equiv T_T(b)$ form a  set of mutually
commuting symmetry operators. In addition to the symmetry operators it is convenient
to define the  energy translation operator \cite{Zak3}
\begin{equation}\label{newop}
{\cal  T}_E \,    =  \exp{ \left(-i\frac{2 \pi }{ \tau_0} t  \right) }   \, , 
\end{equation}
that produces a finite translation in energy by ${2 \pi /  \tau_0} \equiv q b E$.
${\cal  T}_E $ commutes with the three symmetry  operators but not with $S$.
Its eigenfunctions
\begin{equation}\label{quasit}
\mca{T}_E   \psi = e^{iqbE\vartheta} \psi,
\end{equation}     
define a quasitime  $\vartheta$  modulo  $\tau_0$. 

\subsection{Wave function and generalized Bloch conditions}\label{blochcond}

Having defined  ${\cal  T}_0$, ${\cal T}_L$ and
${\cal T}_T$ that commute with each other and  also
with $S$,  it is possible to seek for solutions of the
Schr\"{o}dinger  equation labeled by the quasienergy
(${\cal E}$)  and the longitudinal  ($k_1$) and transverse
($k_2$) quasimomentum according to
\begin{align}
\mca{T}_0 \ket{\mca{E},k_1,k_2} &=e^{-i\tau_0\mca{E}}
\ket{\mca{E},k_1,k_2}, \non
\mca{T}_L\ket{\mca{E},k_1,k_2}  &=e^{ik_1qb}
\ket{\mca{E},k_1,k_2}, \label{trf1} \\
\mca{T}_T \ket{\mca{E},k_1,k_2} &=e^{ ik_2b}
\ket{\mca{E},k_1,k_2}.\num
\end{align}
The magnetic Brillouin zone (MBZ) is defined by 
$k_1\in\yav{0,2\pi/qb}$  and $k_2\in\yav{0,2\pi/b}$.
Similarly the quasienergy  $\mca{E}$
is defined modulo $2\pi/\tau_0=qbE$. If a restricted energy scheme is selected 
for the energy,  the first energy Brillouin region is defined by the condition
$\mca{E}\in\yav{0,2\pi/\tau_0}$.
We shall find convenient to perform a  canonical transformation to new variables
according to
\begin{align}
Q_0&=t,&
P_0&=\mca{O}_0 -\frac{1}{2}E^2,\non
Q_1&=\Pi_2+ E,&
P_1&=\Pi_1 ,\label{ops1}\\
Q_2&=\mca{O}_1 - E t,&
P_2&=\mca{O}_2+E ,\num
\end{align}
that satisfy the commutation rules
$\yav{Q_{\mu},P_{\nu}}=-ig_{\mu\nu}, \quad  g_{\mu\nu}= Diag(-1,1,1)$.
Applied to Eq. (\ref{shr1}) the transformation yields for the \Sh equation 
\begin{equation}\label{shr3}
P_0  \ket{ \psi }=  H  \ket{ \psi}   \, ,  \qquad \qquad 
  H =  \yav{\frac{1}{2} \fues{P_1^2+Q_1^2} +
V\fues{x,y} -E  P_2 } \, .
\end{equation}
In this equation  $(x, y)$ must be expressed in terms of the new variables:
\begin{align}\label{relxQ}
\frac{x }{a} \,=\,  & \frac{m_1 }{ b} \, \left( Q_1 - P_2 \right) - \frac{m_2 }{ b}
\, \left( Q_2 - P_1 \right)  \, , \num \\
\frac{y }{ a} \,=\, & \frac{m_2 }{ b} \, \left( Q_1 - P_2 \right)  + \frac{m_1 }{ b}
\, \left( Q_2 - P_1 \right)  \, .
\end{align}
On the other hand,  the EM-symmetry operators and the 
energy translation operator take the form:
\begin{align}
\mca{T}_0 &=e^{-i\tau_0P_0}, \qquad \qquad 
\mca{T}_L =e^{iqb\fues{Q_2+EQ_0}},\label{cco1}\\
\mca{T}_T &=e^{i bP_2},  \qquad \qquad 
{\cal  T}_E \,   =  e^{ -i \frac{2 \pi }{ \tau_0} Q_0   }   \, .  \num
\end{align}
Notice that these operators do not depend on the variables $P_1, Q_1$. Thus, 
it is  natural to  split the  phase space $(Q_\mu,P_\mu)$ in the  $(Q_1,P_1)$
and the $(Q_0,P_0;Q_2,P_2)$ variables. For the first set   a harmonic oscillator
base is used. Whereas for the subspace generated by the variables
$(Q_0,P_0;Q_2,P_2)$, we observe that the  operators
(${\cal  T}_0^i$, ${\cal T}_L^j$, ${\cal T}_T^k$,  ${\cal T}_E^l$)
with all possible integer values of $(i,j,k,l)$ form a complete
set of operators.
The demonstration follows similar steps as those presented by
Zak in reference \cite{Zak4}.  Hence a complete set of functions,
for the subspace $(Q_0,P_0;Q_2,P_2)$, is provided by the eigenfunctions
of the operators (${\cal  T}_0^i$, ${\cal T}_L^j$, ${\cal T}_T^k$,
${\cal T}_E^l$). As starting point, we select  a base of eigenvalues
of the following operators
\begin{align}\label{bas5}
A^{\dag}A\ket{\mu,\mca{E},k_2}&=\mu\ket{\mu,\mca{E},k_2},\non
P_0\ket{\mu,\mca{E},k_2}&=\mca{E}\ket{\mu,\mca{E},k_2} ,\\
P_2\ket{\mu,\mca{E},k_2}&=k_2\ket{\mu,\mca{E},k_2},\num
\end{align}
where $A $  and $ A^{\dag}$ are the 
lowering and raising operators of  the $\mu$-Landau level:
\begin{equation}\label{raising}
A= \frac{1 }{ \sqrt{2}} (P_1 - iQ_1) \, , \qquad
A^{\dag} = \frac{1 }{ \sqrt{2}} (P_1 + i Q_1) \, . 
\end{equation}
 It is straightforward to verify that these states 
fulfill the required eigenfunction condition (\ref{trf1}) for  the  $\mca{T}_0$
and $\mca{T}_T$ operators. On the other hand $\mca{T}_L$ induces a shift in 
the $\mca{E}$ and $k_2$ labels, while $\mca{T}_E$ induces a shift
$qbE$ in  the $\mca{E}$ label. The previous considerations suggest that a state 
$\ket{\vartheta,\mca{E},k_1,k_2}$ that is a simultaneous  eigenfunction of the
four operators $\mca{T}_E$, $\mca{T}_0$, $\mca{T}_L$ and
$\mca{T}_T$ with the required eigenvalues   (Eqs. \ref{quasit} and \ref{trf1}),
can be constructed as a linear superposition of states of the form
$ \left[ \mca{T}_L \right]^l  \, \left[   \mca{T}_L  \mca{T}_E^{-1} \right]^m %
\ket{\mu,\mca{E},k_2}$.
We write down the state, and verify their correctness: 
\begin{equation}\label{mor3}
\ket{\vartheta,\mca{E},k_1,k_2}=
\sum_l \yav{\mca{T}_Le^{-iqbk_1}}^l
\sum_{\mu,m} c^{\mu}_m\yav{e^{i\sigma bQ_2}
e^{i\sigma b\fues{E\vartheta-k_1}}}^m
\ket{\mu,\mca{E},k_2} \, . 
\end{equation}
It is easy to check that this  function  satisfies the  three  eigenvalue equations
for the symmetry operators  in (\ref{trf1}). 
Additionally it can be verified that  the eigenvalue condition (\ref{quasit}) for
the energy translation operator is also fulfilled 
by  imposing   the periodicity condition $c_{m +p}^\mu =\,  c_m^\mu$.
Every  state in (\ref{mor3}) yields a set of different eigenvalues, a condition that
follows from the fact that 
the selected  set of operators is complete, hence  the base of eigenvalues in
(\ref{mor3})  is complete and orthonormal.   

Up to this point we have constructed a base for the set of four
 (${\cal  T}_0$, ${\cal T}_L$, ${\cal T}_T,{\cal T}_E)$  
operators, however the operator ${\cal T}_E$ is not a symmetry of the
problem, consequently the  solution of the \Sh
is constructed as a superposition of all the states characterized 
by  $\vartheta$;  the  state becomes 
\begin{equation}\label{wav3}
\ket{\mca{E},k_1,k_2}=\int d\vartheta\;
C\fues{\vartheta}
\ket{\vartheta,\mca{E},k_1,k_2}
=\sum_{l} \yav{\mca{T}_Le^{-iqbk_1}}^l
\sum_{\mu,m}b^{\mu}_m
e^{i\sigma b\fues{Q_2-k_1}m}
\ket{\mu,\mca{E},k_2} \, , 
\end{equation}
where
\begin{equation}\label{per4}
b^{\mu}_{m}=\int d\vartheta\;
C\fues{\vartheta}c^{\mu}_m
e^{iqbE\vartheta m} \, . 
\end{equation}
Equation  \rep{wav3}  represents the correct state that satisfies the eigenvalue
equations (\ref{trf1}). It is convenient to  recast it  in a compact form as 
\begin{equation}\label{wav22}
\ket{\mca{E},\bs{k}}=
\mca{W}\fues{k_1} \ket{\mca{E},k_2} ,
\end{equation}
where  $\ket{\mca{E},\bs{k}}=\ket{\mca{E},k_1,k_2}$,
and the operator  $\mca{W}\fues{k_1}$ is defined as 
\begin{equation}\label{wav21}
\mca{W}\fues{k_1}=\sum_{l}\yav{\mca{T}_Le^{-i qbk_1}}^l
=\sum_{l}e^{iqbl\fues{EQ_0+Q_2-k_1}}.
\end{equation}
whereas the ket $\ket{\mca{E},k_2}$ is given by 
\begin{equation}\label{wav25}
\ket{\mca{E},k_2}=\sum_{\mu,m}
e^{i\sigma bm\fues{Q_2-k_1}}b_m^{\mu}
\ket{\mu,\mca{E},k_2}.
\end{equation}
The previous expression stresses the fact that  the quantity
$e^{-i\sigma b m k_1} \, \, b_m^{\mu}$ does not depends on $k_1$,
this will be demonstrated below  Eq. (\ref{har3}). These results will
prove to be very useful to analyze the EM problem, in particular they allow
us to obtain an effective \Sh equation where the ``dynamics" is
governed by the derivative with respect to the longitudinal pseudomomentum.
Utilizing \Eqs{wav22} y \rep{wav21} it is straightforward to prove
the following relation 
\begin{equation}
P_0 \ket{\mca{E},k_1,k_2} = \left[ P_0, \mca{W} \right]  \ket{\mca{E},k_2} +
\mca{W}  P_0  \ket{\mca{E},k_2}
=\fues{\mca{E}- i\bs{E}\cdot\nas{k}}
\ket{\mca{E},k_1,k_2}.
\label{pro2}
\end{equation}
Thus the 
\Sh  (\ref{shr3}) can be recast as 
\begin{equation}\label{pro3}
\fues{\mca{E}-i\bs{E}\cdot\nas{k}}\ket{\mca{E},k_1,k_2}
=H\ket{\mca{E},k_1,k_2} \, . 
\end{equation}
As it will be lately discussed, this form of the Schr\"{o}dinger 
equation becomes very useful to prove various properties of the system, 
in particular  the adiabatic solution of the problem. 

The wave function takes a simply form if we adopt the $(P_0,P_1,P_2)$
representation. In this case
$\Psi_{\bs k }\fues{p}=\left\langle P_0,P_1,P_2\vert {\cal E},%
k_1,k_2  \right\rangle$  yields
\begin{equation}\label{wf2}
\Psi_{\bs{k}} \fues{P}=
\sum_{\mu,l,m} \, b_m^\mu \,  \phi_\mu(P_1) \,
e^{i\left(2 \pi /b \right)  ( m + pl) k_1  }
\delta \left(P_0 - {\cal E} - l q b E \right)
\,\delta\left(P_2- k_{2} + \frac{2\pi}{b}(m + p l ) \right),
\end{equation}
where  $\phi_{\mu}\fues{P_1}$ is the harmonic oscillator function  in the 
$P_1$ representation
\begin{equation}\label{hofun}
\phi_{\mu}\fues{P_1}=\brak{P_1}{\mu}=
\frac{1}{\sqrt{\pi^{1/2}2^{\mu}\mu !}}
e^{-P_1^2/2}H_{\mu}\fues{P_1} \, , 
\end{equation}
and  $H_{\mu}\fues{P_1}$  is the Hermite polynomial.

On the other hand,  it is sometimes useful to restore the space-time  representation 
of the wave function.  This problem provides a good example of the use of canonical
transformations  in quantum mechanics. Identifying the matrices that relate the
$(x_\mu, p_\mu)$ variables to $(Q_\mu, P_\mu)$, it is possible to apply the method of
reference \cite{Moch1} in order to obtain the desired transformation. The final
result for the wave function in the space-time  representation  yields
\begin{equation}\label{flo4}
 \Psi_{ \bs{k}}  \left( t, \bs{x} \right)=
e^{i\bs{k}\cdot \bs{x}-i\mca{E}t} \, \, 
u_{\bs{k}}  \left( t, \bs{x} \right) \, ,
\end{equation}
where the modulation function $u(t,\bs{x}) $ is given by 
\begin{align}\label{wav5}
u_{\bs{k}} \left( t, \bs{x} \right)  =
\frac{1}{\sqrt{2\pi}} & e^{-ix_1 \left[ (\alpha - 1/2) x_2 + k_1 \right]  }
e^{i(\beta +  1/2) E t x_1   }  \nonumber \\
&\times \sum_{\mu l m} i^{\mu}  b_{m}^{\mu}
e^{- iqbElt}
 e^{i\sigma b\fues{k_1 - x_2}\fues{m+pl}}
\phi_{\mu}\fues{x_1- k_2  + 2\pi\frac{m+pl}{b}} \, , 
\end{align}
here  $\phi_{\mu}$ is the same  harmonic oscillator function  \Eq{hofun}, but
now evaluated in the $space$ representation. It is worthwhile to notice, that whereas 
there is not explicit  gauge dependence for the wave function in the  
$P$ representation  (\Eq{wf2}), the dependence 
on the gauge parameters $\alpha$ and $\beta$ becomes  explicit 
in the space-time  representation (\Eq{wav5}).

Solution (\ref{flo4},\ref{wav5}) includes a superposition of Landau-type solutions
originated beneath the spatial and time periodicity.
The spatial periodicity is simply related to the external potential $V$.
Whereas time periodicity arises from the conditions imposed to the symmetry
operators in order to produce commuting symmetries; as discussed in
\Eq{tau1}, the period is given by the time  that takes an  electron
to drift between contiguous lattice points. Notice that Eqs. (\ref{flo4},\ref{wav5})
follows from  the  Bloch and  Floquet  theorem. However in the
electric-magnetic case the modulation functions $u\fues{t, \bs{x}}$
are not strictly periodic, instead they satisfy the 
generalized  Bloch conditions 
\begin{align}\label{flo5}
u(t+\tau_0,x_1,x_2)&=
e^{i\tau_0 \Lambda_0} \, 
u(t,x_1,x_2),\non
u(t,x_1 +qb,x_2)&=
e^{-iqb \Lambda_1} \,
u(t,x_1,x_2),\\
u(t,x_1,x_2+b)&=
e^{-ib \Lambda_2} \, 
u(t,x_1,x_2),\num
\end{align}
the phases are determined by the dual-gauge potential that appears
in the symmetry operators \Eq{dualgauge}. It is straightforward 
to verify that $u_{\bs{k}}$ in \Eq{wav5} satisfies these conditions. 
The generalized Bloch conditions for the magnetic-Bloch problem have been 
previously discussed, our results in (\ref{flo5}) reduces to those in 
reference \cite{Harper3}, when $\bs E =0$ and  the gauge $\alpha=1/2$ is selected. Our results
in (\ref{flo5}) have been  extended in order to include the electric field effects, furthermore  
they apply for an arbitrary  gauge. 
The correct normalization conditions for  the modulation function
are obtained as follows
\begin{equation}
\frac{\fues{2\pi}^2}{qb^2}
\int_{\rm MUC}d^2x\;
u^*\fues{t,x} \, u\fues{t,x}=1 \, , 
\end{equation}
where MUC represents the magnetic cell defined by: 
$x_1\in\yav{0,qb}$ and  $x_2\in\yav{0,b}$.

The function $u$ satisfies conditions similar to those in
\Eq{flo5} but in the MBZ. The MBZ is actually a torus  $T^2$, 
so the edges $(k_1=0,k_2)$ and  $(k_1=2 \pi/ qb,k_2)$ must
be identified as the same set of points, the wave function can differ
at most by a total phase factor (similarly for the edges $(k_1, k_2=0)$
and  $(k_1  , k_2=2 \pi/ b)$):
\begin{equation}\label{flo6}
u(k_1 + 2\pi/ qb, k_2) = e^{if_1} u(k_1,k_2) , \qquad \qquad
u(k_1, k_2+2\pi/ b ) = e^{if_2} u(k_1,k_2),
\end{equation}
the functions $f_1\fuve{k}$ and $f_2\fuve{k}$ can be  related with the quantized 
values of  the Hall conductance \cite{Khomo1}, however instead of \Eq{flo5}, these functions
are not  analytically known and must be computed numerically. 

\section{Harper generalized equation }\label{harper}
\subsection{Finite difference  equation   }\label{finitedif}

The coefficient  $b^{\mu}_{m}$  in Eq.  \rep{wav3} satisfies a recurrence
relation that is obtained when this base is used to calculate the matrix elements
of the  \Sh equation  \rep{shr3}.  First, let us consider the contribution arising
from the  periodic potential  in \Eq{pot0}. 
The $x$ and $y$ coordinates are written 
in terms of the new variables $(Q_1,P_1,Q_2,P_2)$ by means of Eq.  \rep{relxQ},
producing a term of the form   $\exp \left[ (i2\pi r m_1 (Q_1 - P_2)/b \right]$,
and  another contribution  in which $Q_1$ and $P_2$ are replaced by $Q_2$ and $P_1$.
Once that   $Q_1$ and $P_1$ are replaced by  the raising and lowering operators
in \Eq{raising}, one is lead   to evaluate   the matrix elements of the operator
$D=\exp\fues{z A^{\dag} -z ^* A}$ that  generates  coherent Landau states.
A calculation yields
\begin{equation}\label{laguerres}
D^{\nu \mu}\fues{z}=\braket{\nu}{\exp\fues{z A^{\dag}-z^*A}}{\mu}
=e^{-\frac{1}{2}\abs{ z }^2}
\llal{\begin{array}{ll}\fues{-z^{*}}^{\mu-\nu}
\sqrt{\frac{\nu!}{\mu!}}L^{\mu-\nu}_{\nu}
\fues{\abs{z}^2},& \mu >\nu, \\
z^{\nu-\mu}\sqrt{\frac{\mu!}{\nu!}}
L^{\nu-\mu}_{\mu}\fues{\abs{ z}^2},
& \mu <\nu,\\
\end{array}}
\end{equation}
where $L^{\mu}_{\mu}$ are the generalized Laguerre polynomial. 
The  evaluation of the terms that include the $P_2$ operator is direct,
because the base is a  eigenvalue of this operator, on the other hand the
operator $\exp \left[ (i2\pi r m_1 Q_2)/b \right]$ acts as a translation
operator that produces a shift on the index $m$ of the  $b^{\mu}_{m}$ coefficient.
Taking into account these results, it is possible to demonstrate after a lengthly
calculation  that the  \Sh equation  \rep{shr3} becomes 
\begin{equation}\label{idi1}
\sum_{t=-N}^{t=N}  \mb{A}_m^t \fues{k_1,k_2}
{\tilde b} _{m+t}=\fues{\mca{E}+Ek_2+\sigma bEm} {\tilde b}_m \, . 
\end{equation}
Besides its dependence on the index $m$,    ${\tilde b} _m$   is a vector with $L$
components  and  $ \mb{A}_m^t \fues{k_1,k_2}$ is a $L \times L$ matrix   in the
Landau space,    according to 
\begin{align}
{\tilde b_m} &=  \{ b_m^\mu \} =  \fues{b^0_{m},b^1_{m},...,b^L_{m}} \, , \non
\left(  \mb{A}_m^t  \right)^ {\mu \nu} &=
e^{-i\sigma bk_1t}\sum_{r,s}B^ {\mu \nu}_{m}\fues{r,s}
\delta_{t,rm_2-sm_1} \, . 
\label{har2}
\end{align}
In the previous expressions    $L$ is the highest  Landau level   included
in the calculations, $N=max \{ (r,s)\fues{m_1+m_2} \}$ is related to the
largest harmonic  in the Fourier decomposition of the periodic potential
\rep{pot0}, and $B^{\nu \mu}_{m}\fues{r,s}$ is given by 
\begin{equation}\label{deb1}
B_{m}^{\nu \mu}\fues{r,s}=
\llal{\begin{array}{ll}
\fues{v_{00}+\mu+\frac{1}{2}
}\delta^{\mu \nu},& r,s=0,\\
 & \\ v_{rs}D^{\nu \mu}\fues{H_{rs}} e^{iK_{rs}}
e^{iM_{rs}\yav{\sigma b\fues{rm_2-sm_1+m}+k_2}}&r,s\ne0 , 
\end{array}}
\end{equation}
the following definitions where introduced in the previous equation
\begin{align}\label{deb2}
H_{rs}&=\frac{\sqrt{2}\pi}{b}\fues{m_2-im_1}
\fues{ir+s},\non
K_{rs}&=\frac{2\pi^2}{b^2}\fues{r m_2-s m_1}
\fues{r m_1+s m_2},\\
M_{rs}&=-\frac{2\pi}{b}\fues{r m_1+s m_2}.\num
\end{align}
We recall that the indexes $(r,l)$ refer to the Fourier expansion of the periodic
potential \Eq{pot0}, whereas $(m_1,m_2)$ correspond  to the integers that
determined the electric field orientation in  \Eq{rho1}. 

Equation  \rep{idi1} describes the system in terms of a recurrence relation on a
two dimensional  $(\mu, m)$ basis.
The coupling in $m$ ranges from $m - N$ to $m + N$.
We follow the method of reference \cite{Risken1}  to recast  the recurrence
relation on $m$ with $N$ nearest-neighbor coupling into a tridiagonal vector
recurrence relation. This is enforced by defining a $N$ component vector 
\begin{equation}\label{relcb}
{\tilde  c}_m=\fues{\begin{array}{ccccc}
{\tilde b }_{Nm-1}, & {\tilde b }_{Nm}, & {\tilde b }_{Nm+1}, & \dots \dots , 
& {\tilde b }_{N\fues{m+1}-1}
\end{array}} \, , 
\end{equation}
and matrices   $Q^\pm_m$ and $Q_m$  with elements 
\begin{align}
Q^-_m\fues{k_1,k_2}&=\fues{\begin{array}{cccc}
 \mb{A}^{-N}_{Nm} &  \mb{A}^{-N+1}_{Nm} & \dots &   \mb{A}^{-1}_{Nm} \\
0 &  \mb{A}^{-N}_{Nm+1} & \dots &  \mb{A}^{-2}_{Nm+1} \\
\vdots & \vdots & \ddots & \vdots \\
0 & 0 & \dots &  \mb{A}^{-N}_{N\fues{m+1}-1} \non 
\end{array}} \, , \\ \nonumber  \\
Q_m\fues{k_1,k_2}&=\fues{\begin{array}{ccccc}
\mb{A}^{0}_{Nm} &  \mb{A}^{1}_{Nm} & \dots &  \mb{A}^{N-1}_{Nm} \\
\mb{A}^{-1}_{Nm+1} &  \mb{A}^{0}_{Nm+1} & \dots &   \mb{A}^{N-2}_{Nm+1} \\
\vdots & \vdots & \ddots & \vdots  \\
\mb{A}^{1-N}_{N\fues{m+1}-1} &  \mb{A}^{2-N}_{N\fues{m+1}-1} & \dots&
\mb{A}^{0}_{N\fues{m+1}-1}\\
\end{array}}\label{deq1} \, . 
\end{align}
Using the fact that  $Q^+_m = (Q^-_m)^\dag$,  \Eq{idi1} can be reorganized as a
tridiagonal recurrence relation 
\begin{equation}
Q^-_m {\tilde c}_{m-1}+ Q_m {\tilde c}_m+Q^+_m {\tilde c}_{m+1}=
\yav{\fues{\mca{E}+Ek_2+\sigma bEm}I_N+\sigma bED_N} {\tilde c}_m \, . 
\label{har3}
\end{equation}
Here, ${\tilde c}_m$ is $N-$dimensional vector according to the definition
\rep{relcb}, additionally each one  of its   components is a $L$
dimensional vector due to its dependence on the Landau index \Eq{har2}.
$I_N$ is the unit matrix and $D_N=Diag \,\, (0,1,2,\dots,N-1)$.
Likewise,  each of the components of the matrices  $Q_m$,  $I_N$,
and  $D_N$ are  $L \times L$ matrices relative to the Landau contributions.
In relation with \Eq{wav25} it was mentioned that the quantity
$ e^{-i\sigma b  k_1} \, \,  {\tilde c}_m$ does not explicitly depend
on  the longitudinal  pseudomomentum  component $k_1$, this can be easily
verified if the substitution
${\tilde c}_m^\prime  = e^{-i\sigma b  k_1} \, \,  {\tilde c}_m$
is incorporated into \Eq{har3}, then using \Eqs{har2} and \rep{deq1},
it  is straightforward to verify that ${\tilde c}_m^\prime$ is indeed
independent of $k_1$.

\Eq{har3} is one of our main results,  and deserves to be emphasized.
It is a generalization of the Harper equation that exactly  includes
the following  effects:
(1)  an arbitrary periodic potential,
(2) an electric field orientated in a commensurable direction of the
lattice,  and
(3) the coupling between different Landau levels.
So far, no approximations are involved, consequently it holds under
most general conditions. In practice, only a finite number $L$ of Landau
levels can be included on the calculations, but a very good convergence can
be obtained with a reasonable small selection for $L$.
Previous known results are recovered if some approximations are enforced:
($i$) if the electric field is switched off, its orientation is meaningless,
so the integers  $(m_1,m_2)$ can be set to $m_1=1$ and $m_2=0$, in this case
\Eq{har3} reduces to the model previously discussed by
Petschel and Geisel \cite{Geisel1}.
($ii$) If $\bs E = 0$ and additionally the coupling between different
Landau bands  is neglected and the potential is taken as the sum of
cosines given in \Eq{pot1}, then the  system  reduces to a set of Harper
equations (\ref{harp0}), one for each Landau band.
 
Let us return to the case that includes the electric field. Equation \rep{har3}
describes the system dynamics under  most generals conditions; in spite of its
complicated structure the equation can be solved by the matrix continued
fraction methods. However,  a considerable simplification is obtained if  the  electric field is aligned 
along the $x-$lattice axis, i.e. $(m_1=1,m_2=0)$. 
Henceforth, we  consider that the electric field is orientated along the
$x-$axis $(m_1=1,m_2=0)$,  so $b \equiv a$ and additionally  that the 
periodic potential takes the form  given in  \rep{pot1}. The dimension $N$
of the vectors ${\tilde c}_m$  and matrices 
in Eqs. \rep{relcb} and \rep{deq1}  reduces to  $N= max \{ (r,s) (m_1 + m_2\}=1$,
hence the dimensions of  ${\tilde c}_m \equiv c_m$ and   $Q$   reduce  to include
only   the Landau  indexes. Using equations \rep{har2}, \rep{deb1} and \rep{deq1}
the matrices $Q$ become
\begin{align}\label{defQ}
( Q^{+}_m )^{\mu\nu} & \equiv ( Q^{+}  )^{\mu\nu}
=\frac{\lambda \pi K}{2\fues{1+\lambda}a^2} \, e^{\imath\sigma bk_1}
D^{\mu\nu}\fues{\sqrt{\pi \sigma}},\non
(Q_m )^{\mu\nu} &=\frac{\pi K}{2\fues{1+\lambda}a^2} \, 
e^{i\fues{2\pi\sigma m+\sigma bk_2}}
D^{\mu\nu} \fues{\sqrt{\pi \sigma}}+c.c. ,
\end{align}
here the parameter $K$ is a measure of the strength of the coupling
between Landau bands 
\begin{eqnarray}\label{kkk1}
K=\frac{ma^2U_0\fues{1+\lambda}}{\hbar^2\pi}.
\end{eqnarray}
Taking into account these simplifications and that $D_N$ cancels out, we find
that  \Eq{har3} reduces to 
\begin{equation}\label{har5}
Q^- c_{m-1}+ Q_m c_m+Q^+ c_{m+1}=
\fues{\mca{E}+Ek_2+\sigma bEm}c_m.
\end{equation}

%
\subsection{Numerical results}\label{numeric}
The generalized Harper equation is given by a tridiagonal infinite
recurrence relation \rep{har5}. If $\bs E$ is switched off, the equation becomes
periodic with period $p$, in this case the equation can be recast as a finite
$p L \times p L$ matrix that can be solved by a direct diagonalization.
However,  the introduction of $\bs E$ breaks the periodicity,  and most of the
methods used to solve the Harper equation
break down.     \Eq{har5} was solved  
using  an expansion of   the associated Green's
function into matrix continued fractions (MCFs) \cite{Risken1}. 
The energy spectrum is determined detecting the
change of sign in the Green's functions that appear in the vicinity
of a pole. The density of states can be obtained from
the expression 
\begin{equation}\label{gre5}
N\fues{\mca{E}}=
\frac{\imath}{\pi}\mathrm{Tr}\;
\tilde{G}\fues{\mca{E}}=
\mp\frac{1}{\pi}
\mathrm{Im}\yav{\mathrm{Tr}
\;G^{\pm}\fues{\mca{E}}} \, , 
\end{equation}
where the discontinuity in  Green's function is given by 
$\tilde{G}\fues{\mca{E}} =G^{+}\fues{\mca{E}} -G^{-}\fues{\mca{E}}$, 
and the retarded and advanced Green's functions are defined on the side limits 
$G^{\pm}\fues{\mca{E}} =\lim_{\epsilon \rightarrow 0}
G\fues{\mca{E} \pm i\epsilon}$. 
The numerical solution is obtained by truncating the iteration of the (MCFs) after the 
$M-$th term. 
The solution converges if $M$ is large  enough, in the calculations
we observe that in order to obtain a convergence with a precision
of one part in $10^7$, the cutoff can be selected  as
$M \approx 100$.

\begin{figure} [hbt]
\begin{center}
\includegraphics[width=2.5in]{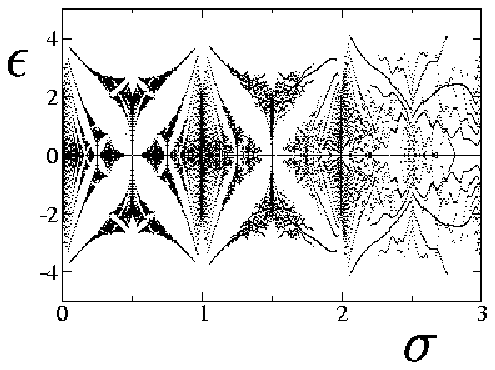}
\end{center}
\caption{The energy spectrum inside the lowest Landau level as a function
of the inverse magnetic flux $\sigma$. The energy  axis is rescaled to
$\epsilon={\cal E} /\left[U_0 e^{-\pi \sigma/2} \right]$. All values of
$k_1\in\yav{0,2\pi/qb}$  are included.
The parameters selected are:
$m^* = 0.067 m_e$, $a = 100 \, nm$, $U_0 = 0.5 meV$,
$E = 0.5 V/cm$, and $k_2 = \pi/2$. }
\label{figure1}
\end{figure}

In our calculations we have used the effective mass $m^* = 0.067 m_e$ typical
for electrons in $GaAs$ and a superlattice
constant   $a = 100 \, nm$ \cite{Klit2}. The rescaled  energy spectrum
${\cal E} /\left[U_0 e^{-\pi \sigma/2} \right]$ is
shown in \Fig{figure1} for the lowest Landau level,
a weak modulation is considered ($U_0 = 0.5 meV$) so
the $\mu-\nu$ Landau mixing is negligible.
The electric field intensity is $E = 0.5 V/cm$ corresponding
to a ratio of the electric to the  periodic potential of
$\rho = e a E/U_0 = 0.02$.   In the strong magnetic region,
$\sigma\in\yav{0,1}$, the  Hofstadter butterfly is clearly depicted.
A distorted replica of the   butterfly spectrum can be still observed
in the region $\sigma\in\yav{1,2}$.
As the magnetic field loses intensity,  the effect of the electric field
becomes dominant, the butterfly is replaced by  discreet levels
separated in  regular intervals, i.e. a Stark ladder. 
Although, $\rho = e a E/U_0$ is small, there is a regime in which the
electric field dominates over the lattice potential  contribution, we can trace
down the origin of this effect to  \Eqs{laguerres} and \rep{defQ}
showing that the periodic contribution is modulated by a factor
$e^{-\pi \sigma/2}$.
Consequently, in addition to a small Landau mixing, the condition
to preserve the butterfly spectrum  can be stated as
\begin{equation}\label{bound1}
e a E \ll U_0 e^{-\pi \sigma/2} \, .
\end{equation}
 This condition  restricts
the intensity of both the electric and magnetic fields,  these are estimated as
(inserting units):  $E \ll U_0/(e a) \sim 100 \, (U_0/meV) \, V/cm$  and
$B \geq \pi h /(e  a^2) \sim 0.6 T $.

\begin{figure} [hbt]
\begin{center}
\includegraphics[width=4.0in]{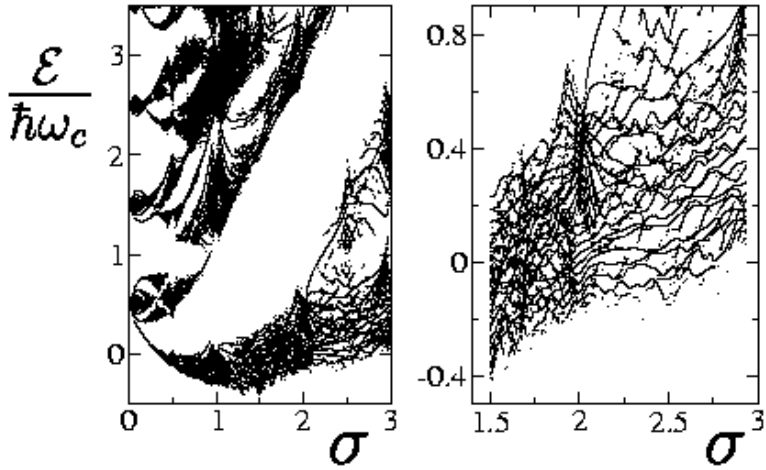} 
\end{center}
\caption{The energy spectrum for the four lowest Landau bands is plotted
as a function of  
magnetic flux $\sigma$. The  coupling strength $K =6$  yields important
Landau mixing. 
All values of $k_1\in\yav{0,2\pi/qb}$  are included. The parameters selected are:
$m^* = 0.067 m_e$, $a = 100 \, nm$, $U_0 = 1.2 meV$, $E = 0.5 V/cm$,
and $k_2 = \pi/2$. The right inset shows  the details of the spectrum 
in a reduced region that was originated from the lowest Landau band. Seven  Landau levels are included in the calculation
in order to attain convergence.}
\label{figure2}
\end{figure}

 \Fig{figure2} shows the energy spectrum for the four lowest Landau bands 
as a function of $\sigma$. In this case  a higher coupling strength ($U_0 = 1.2 meV$)  yields important 
Landau mixing. For a strong magnetic field  the electric field contribution  is
small and the spectrum is very similar to that previously obtained by Petschel
and Geisel \cite{Geisel1}. Three regimes can be identified in the plot:
$(i)$ In the  strong magnetic field limit (small $\sigma$), the Landau bands are well
separated and re-scaled  butterfly structures are clearly observed.
$(ii)$ In the intermediate region ($\sigma  \sim 1$) the  periodic potential
induces Landau level overlapping,
$(iii)$ as $\sigma$ increase  very narrow levels develop as a result of the
electric field, this effect is clearly depicted in the right inset of  \Fig{figure2}.

\begin{figure} [hbt]
\begin{center}
\includegraphics[width=4.0in]{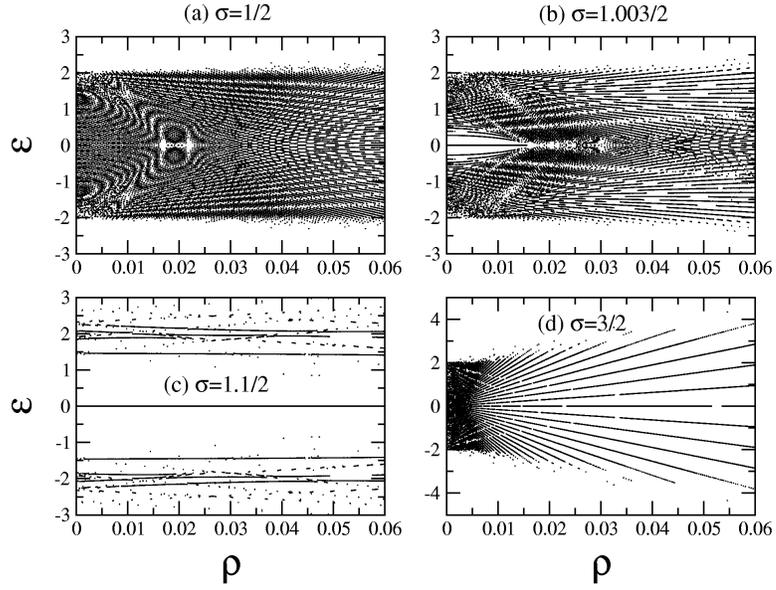} 
\end{center}
\caption{The energy spectrum for a single Landau level 
as a function of the  electric field intensity ($\rho = e a E/U_0$) for:
(a) $\sigma=\frac{1}{2}$, (b) $\sigma=\frac{1.0038}{2}$, (c) $\sigma=\frac{1.1}{2}$,
(d) $\sigma=\frac{3}{2}$.
 The energy  axis is rescaled to
$\epsilon={\cal E} /\left[U_0 e^{-\pi \sigma/2} \right]$.
 All values of $k_1\in\yav{0,2\pi/qb}$  are included, and the other parameters 
 are selected as:
$m^* = 0.067 m_e$, $a = 100 \, nm$, $U_0 = 0.5 meV$,
 and $k_2 = \pi/2$.}
\label{figure3}
\end{figure}

\begin{figure} [hbt]
\begin{center}
\includegraphics[width=2.5in]{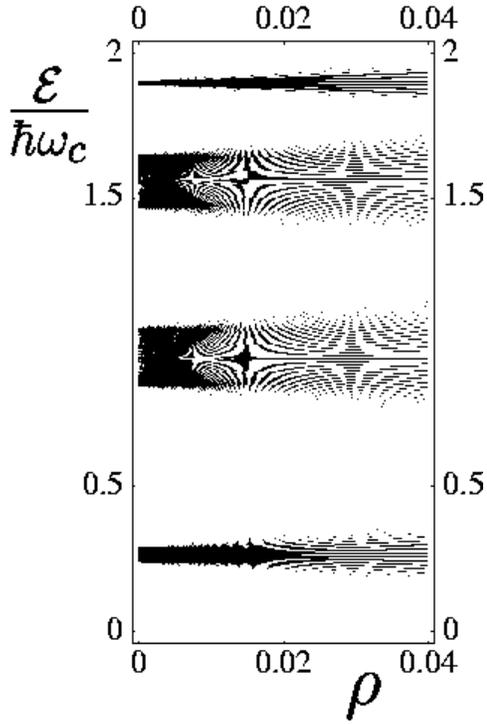}
\end{center}
\caption{Energy density plot  for the two  lowest Landau levels as a function
of the  electric field intensity ($\rho = e a E/U_0$) for 
$\sigma=1/2$. All values of
$k_1\in\yav{0,2\pi/qb}$  are included, the other parameters take the same values as in \Fig{figure2}. }
\label{figure4}
\end{figure}

The effects produced by the electric field can be clearly identified in 
\Fig{figure3} and \Fig{figure4}, where the energy spectrum  as a function of the 
 electric field intensity ($\rho = e a E/U_0$) are displayed. In the first case 
 (\Fig{figure3}) Landau mixing is neglected and the results for a single band
 and various selections of $\sigma$ are displayed. We first notice that for $\sigma =1/2$ 
 and $\bs E =0$ the two band are degenerated at ${\cal E}=0$, this degeneracy
 is lifted for  $\sigma =1.0038/2$; a further increment of  $\sigma$ to
  $\sigma =1.1/2$ yields a transformation of the continuous  bands into  a series of  
  quasi discreet levels. For  $\sigma =3/2$  the band structure is observed at small 
  values of $\bs E$; however as the electric field intensity is increased a transition from a continuos band
  to a discreet level spectrum is  observed.     
 The corresponding plot in  \Fig{figure4} includes Landau mixing and shows  the energy spectrum as
a function of ($\rho = e a E/U_0$) for the first two Landau bands.
For $\sigma =1 /2$ every Landau level splits in two bands, the degeneracy is lifted by the finite Landau mixing.
Again as $E$ increases we observe that 
these bands evolve into  a series of quasi discreet levels.
In these plots we observe that the evolution from the band to discreet structure of 
the spectrum produces a interesting pattern in the distribution of  gap regions, this structure can be related to 
a series of avoided and exact crossing effects of the quasienergies; these effects become more  evident in \Fig{figure3}c. 
It was  pointed out in reference  \cite{Grosmann} that the 
coexistence  of  avoided and exact crossing of quasienergies in the  Floquet spectrum can be understood
on the basis of a generalized parity symmetry; in such a way that  states belonging 
to the same parity number develop avoided crossing, but those of different 
parity symmetry can yield exact crossing.

\begin{figure} [hbt]
\begin{center}
\includegraphics[width=4.0in]{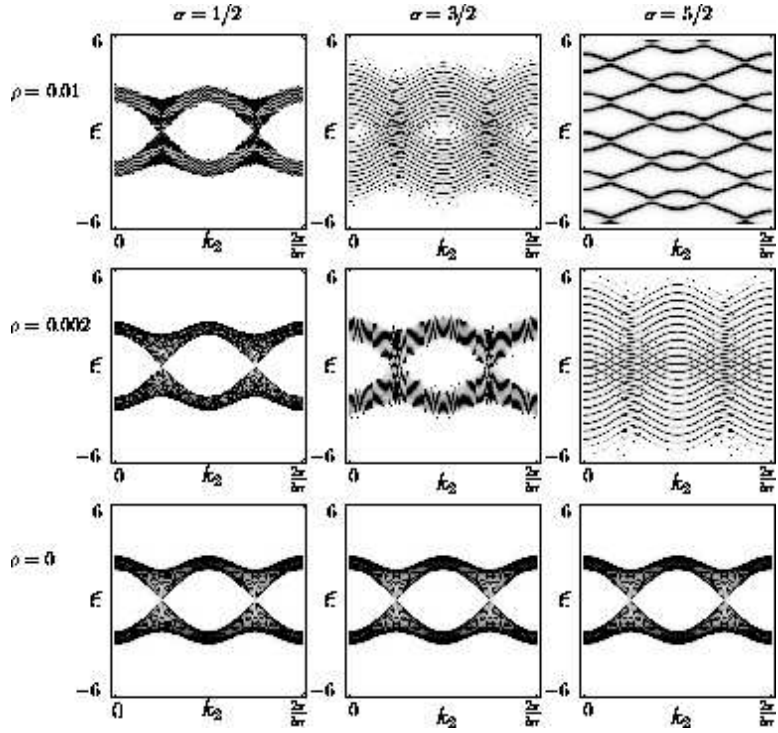} 
\end{center}
\caption{Energy density plot  for the  lowest Landau levels as a function
of the  transversal pseudomomentum $k_2$. Three values of $\sigma=1/2,3/2,5/2$
are selected, so  two bands  appear when  $E = 0$. Three values of $E$
are considered: $\rho = e a E/U_0= 0, \,0.002,\, 0.01.$ }
\label{figure5}
\end{figure}

In \Fig{figure5} the density of states
as a function of the $k_2$ pseudomomentum and the quasienergy  ${\cal E}$
is displayed. The density of states
was calculated from  Green's function discontinuity obtained from the
continued fraction expansion. 
Three values of $\sigma=1/2,3/2,5/2$ are selected, consequently   two bands
appear when  $\bs E = 0$. 
In the superior panels it is observed that both  as $E$ increases or $B$ 
decreases, the spectrum for a fixed value of $k_2$ evolves  into a set of
narrow bands. When  $k_2$ is varied the spectrum becomes almost continuous,
except for the small gaps that open between the bands. 
 
Based on the  previous results it follows  that the EMB model preserves the
band  structure. For weak  electric fields the bands are
grouped forming the ``butterfly  spectrum"; whereas as the intensity of $\bs E$
increases the bands are replaced by a  series of quasi discreet levels:
a ``magnetic a Stark ladder".

\subsection{Adiabatic approximation}\label{adiabatic}
In this section we exhibit an alternative expression for the effective \Sh
equation that governs the dynamics of the  system. Based on  this formalism
we shall  find an  approximated adiabatic solution, that  throws  further
insight into the physical results obtained in the previous  section. Let us
consider the generalized Harper equation \rep{har5},  taking into account
that the matrix $Q_m$ is periodic in  $m$ with period $p$, we find convenient
to define the unitary transformation 
\begin{equation}\label{dia1}
d_l (\phi)  = \sum_m \mb{U}_{l,m}(\phi)  \, c_m \, ,  \qquad
\qquad    \mb{U}_{l,m}\fues{\phi}=\sqrt{\frac{q a}{2\pi}}
e^{iq a \phi  [ m/p ]} \, \delta_{l, m mod[p]}  \, ,
\end{equation}
where $ [ m/p ]$ denotes the integer part of the number, and the delta enforces
the condition $ l = m, \, mod[p]$. Notice that whereas the index $m$ in $c_m$
runs over all the integers, the corresponding label of the state $d_l (\phi)$
takes the values $l =0,1,2, \dots ,p-1$. The new vector state $d_l (\phi)$
satisfies the periodicity condition 
\begin{equation}\label{periodi10}
d_l \left(\phi + \frac{2 \pi }{ q a} \right) \, = \, d_l (\phi) \,, 
\end{equation}
 and the transformation matrices  fulfill  the following properties 
\begin{align}
\sum_{m=-\infty}^\infty  \mb{U}_{l,m}\fues{\varphi}
\mb{U}_{m,l^\prime}^{\dag}\fues{\phi}&=
\delta_{l,l^\prime} \, \delt{\phi-\varphi},  \nonumber\\
\sum_{l=0}^{p-1} \int_{0}^{\frac{2\pi}{q a}}d\phi\;
\mb{U}_{m,l}^{\dag}\fues{\phi}\mb{U}_{l,m^\prime}\fues{\phi} &
=\delta_{m,m^\prime} .
\label{nor1}
\end{align}
Applying  this transformation to  the generalized  Harper equation
\rep{har5} yields
\begin{equation}\label{hep10}
H_M\fues{k_1+\phi,k_2}
d\fues{k_1+\phi,k_2}=\fues{\mca{E}+Ek_2
-iE\frac{\partial}{\partial \phi}}
d\fues{k_1+\phi,k_2},
\end{equation}
where the  Hamiltonian is   reduced to a   $p  \times p$  block  form 
\begin{equation}\label{hep16}
H_M\fues{k_1+\phi,k_2}=
\fues{\begin{array}{cccccccc}
  Q_0    &  Q^+   &     0     &   0       &  \dots  & Q^-  \\
 Q^-   &   Q_1    &   Q^+   &   0       &  \dots  &   0    \\
   0     &  Q^-   &   Q_2     &   Q^+   &  \dots  &   0    \\
 \vdots  &  \vdots  &  \vdots   &  \vdots   &  \ddots & \vdots \\ 
Q^+&    0     &     0     &    0      &  \dots  &Q_{p-1} \\
\end{array}}
\end{equation}
here  $Q^- \equiv Q^- \fues{k_1+\phi,k_2}$,
$Q_m\equiv Q_m\fues{k_1+\phi,k_2}$ and
$Q^+ \equiv Q^+ \fues{k_1+\phi,k_2}$.
Taking into account that each block in \rep{hep16} is a $L \times L$ matrix
determined by the Landau indexes,
we have obtained that  the transformation \rep{dia1}  reduces the infinite
dimensional representation of the \Sh equation to a $pL \times  pL$ representation.
The price is that   a non-local derivative term respect to  parameter $\phi$
has been added.  In the absence of electric field, \Eq{hep10}
represents a finite dimensional   eigenvalue problem, that can be solved by direct
diagonalization. We observe  from Eqs.  \rep{hep10} and \rep{hep16} that
$\phi$ appears to be directly  related the longitudinal quasimomentum  $k_1$,
in fact if we redefine $k_1+\phi\rightarrow k_1$  in \Eq{hep10} yields 
\begin{equation}\label{hep14}
H_M \fues{k_1,k_2}d\fues{k_1,k_2}=\fues{\mca{E}+Ek_2
-iE\frac{\partial}{\partial k_1}}d\fues{k_1,k_2}.
\end{equation}
This expression confirms the form of the  \Sh
equation  \rep{pro3} previously discussed, in which  the ``dynamics" is determined 
by a differential equation with respect 
to the longitudinal quasimomentum. However we now have an 
explicit  finite-dimensional matrix representation for the Hamiltonian.

Let the  ``instantaneous" eigenstates of $H_M$  be $ h^{(\alpha)}$ with
energies $\Delta^{(\alpha)}$, i.e.
\begin{equation}\label{ada0}
H_M   h^{(\alpha)} \left(k_1,k_2\right)=
\Delta^{(\alpha)}  \left(k_1,k_2\right)   h ^{(\alpha)} \left(k_1,k_2\right) \, , 
\end{equation}
where $\alpha$  labels  the band state of the non-perturbed problem.
A solution of  Eq. (\ref{hep14}) can  readily be obtained in the adiabatic
approximation as follows
\begin{equation}
  d^{(\alpha)} \left(k_1,k_2\right)=\exp\left[-i\frac{\cal E}{E} k_1
+i\frac{1}{E}\int_{0}^{k_1}d\phi \,\,  \Delta^{(\alpha)} \left(
\phi , k_2 \right)+i\gamma^{(\alpha)} \left(k_1,k_2\right)\right] 
 h^{(\alpha)} \left(k_1,k_2\right) \, ,
\label{ada1}
\end{equation}
where the Berry phase  $\gamma\left(\phi\right)$ is determined from the substitution 
of the previous  expression in Eq. (\ref{hep14})
\begin{equation}
\gamma^{(\alpha)} \left(k_1,k_2\right)= \,i \int_0^{k_1}d\phi  \,
h^{(\alpha) \dagger} \left(\phi,k_2\right)    
\frac{\partial}{\partial \phi}  h^{(\alpha)}  \left(\phi,k_2\right) \, . 
\label{berry}
\end{equation}
The energy eigenvalue is then determined by the periodicity
condition \rep{periodi10},  hence the change of the phase of the wave
function \rep{ada1} must be an integral multiple of $2\pi$, consequently
the  spectrum in the  adiabatic approximation is given as 
\begin{equation} \label{enersp} 
\mca{E}^{(\alpha)} \fues{k_2}= n E a \sigma
+\frac{q a }{2\pi}\int_{0}^{2\pi/q a}dk_1 \Delta^{\alpha}
\fues{k_1,k_2}\\
+\frac{q a E}{2\pi} \, \gamma^{(\alpha)}(2\pi/q a, k_2)
, \quad n=0,\pm 1, \dots
\end{equation}

This result deserves some comments. The energy spectrum in the presence of
electric and magnetic field contains a series of discreet levels separated by 
\begin{equation}
\Delta \mca{E} =  E a \sigma = \frac{ h E }{ a B},
\label{stark}
\end{equation}
where we have used \Eq{sig1} and restored units. These levels are similar to the Wannier levels that
appear when an electric field is applied to an electron in a periodic potential,
the energy separation being proportional to   $a E$.
In the present  case, the band structure parallel to the electric field is
replaced by a set of discreet steps, this ``magnetic Stark ladder"
are characterized by a separation proportional to the electric field intensity,
but inversely proportional to both the lattice separation  and the magnetic field. 
The  existence of these levels can be explained by the following argument.
In the presence of simultaneous $\bs E$ and $B$ fields,
the electron travels  between lattice points in a time $\tau = a B /E$.
As long as the electron does not tunnel into another band, the motion appears
as periodic, with frequency $\omega = 2\pi / \tau$,   corresponding to a series
of energy levels whose separation $\Delta \mca{E}  = \hbar \omega$ coincides with the
result in \Eq{stark}. This magnetic-Stark ladder is combined with the Hofstadter
spectrum represented by the second term in \Eq{enersp} and the contribution of
the Berry phase, the competition between these three factors was  discussed in
the numerical analysis of the previous  section.  We notice that  the Berry phase
is written as the integral of the longitudinal component of the Berry connection
$\mca{A}\sa\fuve{k}$ discussed by Kohmoto \cite{Khomo1}.


\section{Summary} 
We have considered the quantum mechanics of a periodic electron system 
driven by electric and magnetic fields.
We presented a thoroughly discussion of the  symmetries of the electric-magnetic Bloch problem.
These symmetries were utilized in order to construct the wave function of the system. 
The space-time representation of the wave function  (\ref{flo4},\ref{wav5})
fulfills the Bloch-Floquet theorem  with modulation functions that satisfy 
the generalized Bloch conditions given in (\ref{flo5}); the  phases  that appears in these conditions are determined by the 
dual-gauge potential related to  the symmetry  operators (\ref{dualgauge}).

The dynamics of the system is
governed  by   a finite difference equation (\ref{har3}), that represents a generalization of the Harper equation,
the equation   includes the following  effects:
(1)  an arbitrary periodic potential,
(2) an electric field orientated in a commensurable direction of the
lattice,  and
(3) the coupling between different Landau levels.
The model previously discussed by Petschel and Geisel \cite{Geisel1}, as well 
as the  Harper equation (\ref{harp0}) are recovered if the appropriated limit is enforced.

A detailed numerical analysis based on the solution via a matrix continued fractions  was carried out, and
the properties of the generalized  Hofstadter spectrum were discussed. 
In the  strong magnetic field limit  the Landau bands are well
separated and re-scaled  butterfly structures emerge; 
 in the intermediate region  the  periodic potential
induces important Landau level overlapping; finally as 
  the electric field intensity is increased a transition from a continuos band
  to a discreet level spectrum is  observed. In this ``magnetic Stark ladder" the 
equally spaced discreet levels are separated in  integer multiple of  $ h E / a B $.  
We find that in  order to preserve the self-similarity structure of the Butterfly spectrum, besides the strong magnetic field condition 
$B \geq \pi h /(e  a^2) $,
the electric field  intensity should be  restricted according to \Eq{bound1}.

An effective equation of motion in  $k-$space was derived (\ref{hep14}).
In this equation the   ``dynamics" is governed  
by a differential equation with respect 
to the longitudinal quasimomentum, the  explicit matrix representation for the Hamiltonian is 
given in (\ref{hep16}). Based on this formalism an approximated adiabatic solution for the 
energy spectrum  (\ref{enersp})
was obtained.  This expression  includes: (1) the discreet  ``magnetic Stark ladder" term (\ref{stark}), 
(2) and averaged Hosftadter contribution, and (3) the Berry phase contribution (\ref{berry}). This Berry phase 
is written as the integral of longitudinal component of the 
Berry connection  discussed by  Kohmoto \cite{Khomo1} in relation with the quantization of the Hall conductance. 
We finally remark that the present formalism should set the basis
for the study of the Hall conductivity beyond the linear response 
approximation. 
\vskip 0.5 cm
We acknowledge the partial financial support endowed by
CONACyT through grant No. {\it 42026-F}.

\bibliographystyle{elsart-num}
\bibliography{article}

\newpage

\end{document}